\documentclass[reprint,amsmath,amssymb,aps]{revtex4-2}
\usepackage{graphicx}
\usepackage{dcolumn}
\usepackage{braket}
\usepackage{bm}
\usepackage{subfigure}
\usepackage{algorithm}  
\usepackage{algorithmicx}  
\usepackage{algpseudocode}  
\usepackage{amsmath}  

\begin{document}
	
\preprint{APS/123-QED}
	
\title{Evolution of phase transition in the finite-fugacity extended dimer model}
	
\author{Hongxu Yao}\thanks{These authors contributed equally.}
\author{Jiaze Li}\thanks{These authors contributed equally.}
\author{Jintao Hou}\thanks{These authors contributed equally.}
\author{Jie Lou}\altaffiliation{Corresponding author:
}
\author{Yan Chen}\altaffiliation{Corresponding author: yanchen99@fudan.edu.cn.
}
\address{Department of Physics and State Key Laboratory of Surface Physics, Fudan University, Shanghai 200438, China}

\date{\today}
	
\begin{abstract}
We investigate the evolution of phase transition of the classical fully compact dimer model on the bipartite square lattice with second nearest bonds at finite temperatures. We use the numeric Monte Carlo method with the directed-loop algorithm to simulate the model. Our results show that the order of the phase transition depends on the fugacity of the second nearest bonds. We find that the phase transition reduces from the Kosterlitz-Thouless transition to unconventional high-order phase transitions which feature the coexistence of properties Kosterliz-Thouless transition and the first-order phases transition simultaneously. As the fugacity increases further, phase transition evolves to the first-order phase transition. In addition, our results of dimer-dimer correlation functions and their corresponding structure factor functions computed by us show the evolution of decay correlation for different fugacity.
		
PACS number(s): 74.20.Mn, 75.10.Jm, 05.10.Ln, 05.50.+q
\end{abstract}

\maketitle

\section{\label{sec:level1}Introduction}

Dimer models are pivot models in strong correlated physics corresponding to high-temperature superconductors with resonating valence bonds (RVBs).\cite{doi:10.1126/science.235.4793.1196,doi:10.1080/14786439808206568,BASKARAN1987973} Rokhsar and Kivelson et al. simplify RVBs as a pair of parallel dimers and introduce quantum dimer model (QDM) Hamiltonian considering quantum fluctuations on the square lattice.\cite{PhysRevB.35.8865,PhysRevLett.61.2376} For QDM, the Hamiltonian covers two parts. The former part is kinetic term, there exits flipping intensity for a pair of parallel dimers. The later part means potential intensity of each resonating pair. From the proposal of QDM to now, properties of quantum critical point at zero temperature of different lattices or higher dimensions are studied by numeric or analytical methods.\cite{PhysRevB.103.094421,PhysRevLett.86.1881,PhysRevB.71.224109,Topologicalphasetransition,PhysRevB.73.094430,PhysRevB.64.144416,PhysRevB.68.184512}
	
If the flipping coupling constant and the potential coupling constant is equal (called RK point), the ground state is that all dimer configurations superpositions with the same weight. The dimer correlations at RK point are similar to the correlations of purely classical dimer model (CDM) computed by Kasteleyn’ s theorem.\cite{PhysRev.132.1411,doi:10.1063/1.1703953,PhysRev.124.1664} This means CDM retains relevance to QDM. In addition, in the limit of neglection of kinetic term, it does not impact on the understanding of finite-temperature phase diagram of QDM. Previously, researchers studied CDMs with interaction on the square lattice and reported the phase transition is a Kosterlitz-Thouless(KT) transition.\cite{PhysRevLett.94.235702,PhysRevE.74.041124} The decay with distance of dimer-dimer correlation as $r^{-2}$ at infinite temperatures and algebraic decay at finite temperatures. The phase transition is from fourfold degenerate and ordered columnar dimer crystals to a Coulomb gas. After this, more researches focus on CDMs for higher dimensions and on different lattices. The phase transition of three-dimensional classical dimer model is a continuous phase transition between an ordered dimer crystal and a Coulomb liquid.\cite{PhysRevLett.91.167004,PhysRevLett.97.030403,PhysRevB.78.100402} On the triangular lattice, the correlations of dimer-dimer and monomer-monomer are exactly computed by Kasteleyn’s theorem.\cite{PhysRevB.66.214513} Numeric, if the triangular lattice is anisotropic, researcher found the evolution from short-range disordered liquid dimer phase to a critical phase, then to an order columnar phase as interaction strength increases. And if for isotropic case, there is a first-order transition to an ordered phase.\cite{PhysRevE.76.041125} 
	
To date, more novel extensions are introduced to CDM by researchers for studying correlation in these phases and the phase transition between them.\cite{PhysRevE.76.041125,PhysRevB.82.014429,PhysRevB.73.144504} Traditionally, CDM requires there is exactly one bond with the nearest point for every point on the lattice, these dimers are called $N1$ dimers. Here, we discuss the finite-temperature phase transition of CDM covering second-nearest-neighbor ($N4$) dimers of different fugacity (called extended classical dimer model, ECDM) on square lattice using an energy-based directed-loop Monte Carlo algorithm. Traditional CDM with nearest-neighbor dimers emerges abundant physics but we still do not understand the physics if we extend geometric conditions. Considering the preservation of bipartite lattice structure of traditional CDM\cite{PhysRevLett.94.235702,PhysRevE.74.041124}, we jump over the next-nearest-neighbor ($N2$) dimers and study $N4$ dimers directly as a step forward. In early research, an arbitrarily small fraction of $N2$ dimers leads to exponential dimer correlations and deconfinement.\cite{PhysRevB.73.144504} However, their results of dimer-dimer correlations in $N1-N4$ dimer model are still algebraic. The decay powers of different fugacity in $N1-N4$ dimer model are different. This means types of finite-temperature phase transitions of different fugacity are dissimilar. Thus, we hope to study the finite-temperature phase transition process of finite-fugacity $N4$ dimer model. Previous researches just compute the correlations of small fraction of $N4$ dimers but we compute larger here. 
	
We now explain the outline of the paper in the following: We describe details of ECDM in Sec. II. We present our simulation results of finite-temperature phase transition in Sec. III. In addition, in Sec. IV, we also discuss the dimer-dimer correlations of different fugacity of $N4$ dimers in ECDM. We conclude with a summary and discuss new perspective of eCDM in Sec. V. And in Appendix A, Our energy-based usage of directed loop algorithm and and detailed balance condition as a development are discussed here.
	
\section{\label{sec:level1}EXTENDED CLASSICAL DIMER MODEL}
	
The model studied by us is close-pocked dimers on the square lattice. In the model, $N1$ dimers and $N4$ dimers are coexistence. $N1$ dimers still interact with parallel dimers in a plaquette. $N4$ dimers just occupy points on the lattice and there is no interaction between them. Thus the partition function and the Hamiltonian can be written as the traditional CDM on the square lattice.\begin{equation}
	Z=\sum_{c}exp[-\frac{k}{T}(N(\mathop{\rule[0.1pt]{0.23cm}{0.05cm}}^{\rule[-0.1pt]{0.23cm}{0.05cm}})+N(\rule[-0.6pt]{0.05cm}{0.23cm}            \kern0.08cm\rule[-0.6pt]{0.05cm}{0.23cm}))]
\end{equation}
\begin{equation}
	H=\sum_{c}V[N(\mathop{\rule[0.1pt]{0.23cm}{0.05cm}}^{\rule[-0.1pt]{0.23cm}{0.05cm}})+N(\rule[-0.6pt]{0.05cm}{0.23cm}            \kern0.08cm\rule[-0.6pt]{0.05cm}{0.23cm})]
\end{equation}
$c$ means coverings fully packed by $N1$ and $N4$ dimers on the square lattice. $T$ is the temperature and $v$ is the potential energy between two interacting dimers. In this paper, we set $V=-1$. Vertical and horizontal dimers degenerate with each other and bipartite lattice provides two equivalent occupations for dimers. Two upper reasons require the ground states of our Hamiltonian are four-fold degenerate columnar states.\cite{PhysRevB.35.8865,PhysRevLett.94.235702,PhysRevE.74.041124}
	
Sandvik et al introduced detailed balance condition to CDM covering $N2$ dimers in ref\cite{PhysRevB.73.144504}. For our simulation, we need to develop their theory and introduce it to our energy-based directed loop algorithm. As the ref\cite{PhysRevB.73.144504}, we suppose there are four weights: $a_{ss}$, $a_{ll}$, $a_{ls}$ and $a_{sl}$. The first subscript stands for types of dimers to be updated and the second subscript stands for a new-born dimer after the update. Subscript $s$ means $N1$ short dimer and subscript $l$ means $N4$ long dimer. There are four directions to form $N1$ dimers and eight directions to form $N4$ dimers. The total weight of two dimers of different types can be computed like the this:
\begin{equation}
	\omega_s=4a_{ss}+8a_{sl}
\end{equation}
\begin{equation}
	\omega_l=8a_{ll}+4a_{ls}
\end{equation}
we define the weight fugacity of $N4$ dimers as the following and the detailed process of the algorithm is illustrated in Appendix A:\begin{equation}
	f=\omega_l/\omega_s
\end{equation}
If $f=0$, the model reduces to traditional CDM. At zero temperature, the model is frozen in columnar dimer crystal and undergoes a KT transition to Coulomb gas when $T$ reaches critical temperature $T_c$\cite{PhysRevLett.94.235702,PhysRevE.74.041124}. The easiest excitation from the ground state is to flip one of a pair of parallel dimers costing $-2V$ energy. If $f>0$, however, $N4$ dimers enter the state gradually as a new type of excitation. The easiest $N4$ excitation is updating a pair of dimers spaced at an interval of 2. and generating two cross-over $N4$ dimers costing $-4V$ energy. The model studied by us is schematically shown in Fig. 1.
	
\begin{figure}[htbp]
	\centering
	\label{1} 
	\includegraphics[scale=0.4]{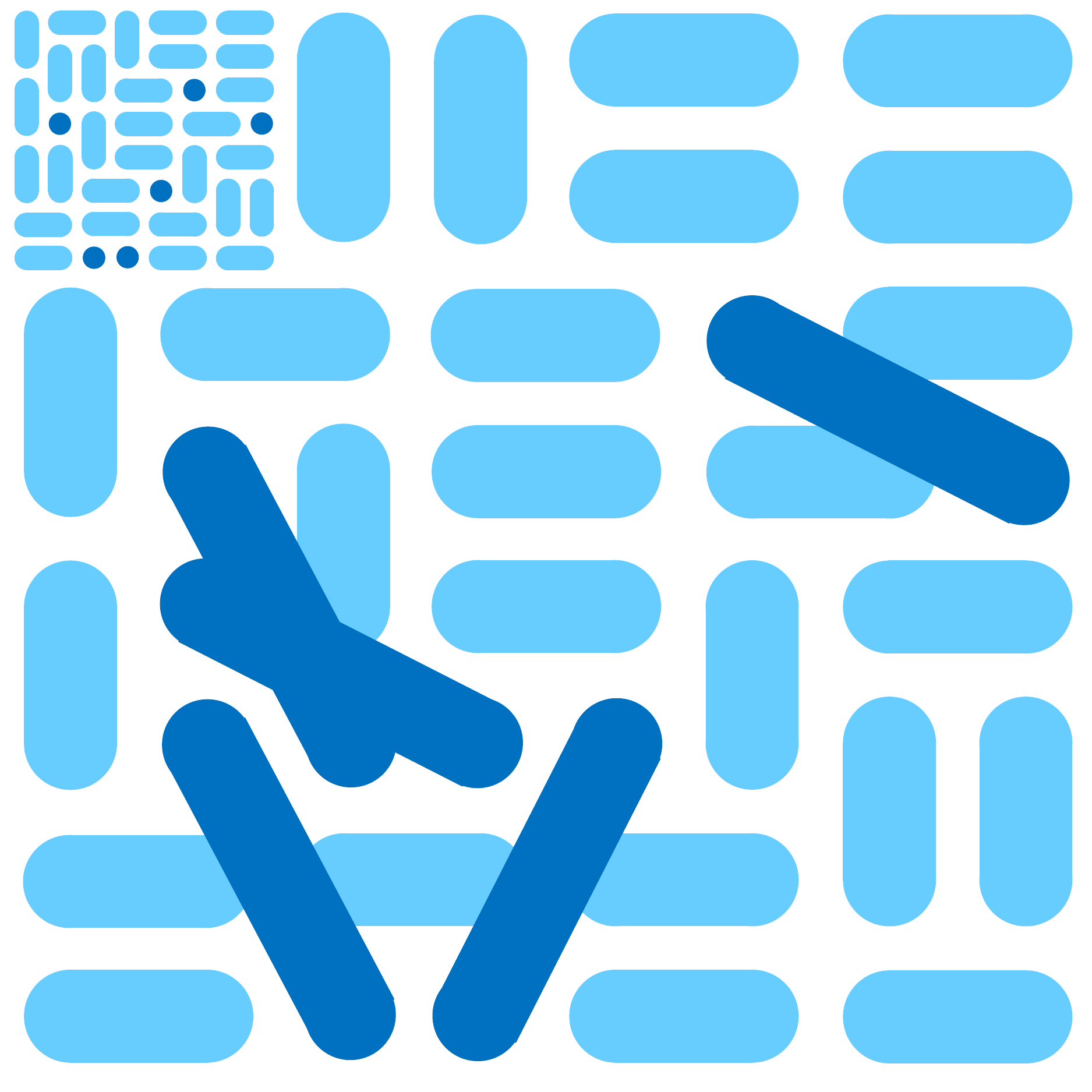}
	\caption{Schematic figure of ECDM. The model extended from traditional CDM owns both $N1$ dimers and $N4$ dimers. Two parallel dimer in a plaquette interact with each other contributes potential energy but $N4$ dimers without interaction and intrinsic energy. Inset: Schematic figure of dimer-hole model introduced to compare the impact for geometric condition of $N4$ dimers and hole defects.
	}
\end{figure}

For a better studying of the impact of $N4$ dimers, we also simulate a dimer-hole model shown in the inset of Fig. 1. We count the average of $N4$ dimers $n_{N4}$ at every temperature, the value $n_{N4}$ means there are $2n_{N4}$ holes in CDM.  The difference between the $N4$ dimers and holes in CDM is that every $N4$ dimer is two bound holes actually. Thus, we introduce free holes of the same value in CDM to study the uniqueness of the geometric condition of $N4$ dimers in the process of phase transition at finite temperatures.
	
Two key points when covering $N4$ dimers in our study should be considered. Firstly, the way to break the symmetry of the ECDM. When $f = 0$, in CDM, order parameters show long-range order in columnar crystal phase but vanish in Coulomb gas at high temperatures shown in ref\cite{PhysRevLett.94.235702} and ref\cite{PhysRevE.74.041124}. Dimer rotational symmetry breaking (DSB) order parameter and pair rotational symmetry breaking (PSB) order parameter describes $\pi/2$-rotational symmetry of dimers.
\begin{equation}
	DSB = N^{-1}|N(\rule[2.5pt]{0.23cm}{0.05cm})-N(\rule[0pt]{0.05cm}{0.23cm})|
\end{equation}
\begin{equation}
	PSB =N^{-1}|N(\mathop{\rule[0.1pt]{0.23cm}{0.05cm}}^{\rule[-0.1pt]{0.23cm}{0.05cm}})-N(\rule[-0.6pt]{0.05cm}{0.23cm}            \kern0.08cm\rule[-0.6pt]{0.05cm}{0.23cm})|
\end{equation}
For bipartite lattice, there is also the translation symmetry. We can define COL as complex columnar order parameter.\begin{equation}
	\begin{aligned}
			COL &= \frac{2}{L^2}|\sum_{\textbf{r}}\psi_{col}(\textbf{r})|\\
			&=\frac{2}{L^2}|\sum_{\textbf{r}}(-1)^{r_{x}}[n(\textbf{r}+\frac{\textbf{x}}{2})
			-n(\textbf{r}-\frac{\textbf{x}}{2})]\\
			&+i(-1)^{r_{y}}[n(\textbf{r}+\frac{\textbf{y}}{2})-n(\textbf{r}-\frac{\textbf{y}}{2})]|
	\end{aligned}
\end{equation}
Where $\textbf{x}$ and $\textbf{y}$ are unit vectors (1,0) and (0,1). We define dimer occupation number as $n$, for example, $n(\textbf{r}+\frac{\textbf{x}}{2})=1$, if there is a dimer between site $\textbf{r}$ and site $\textbf{r}+\textbf{x}$. 
	
Secondly, at infinite temperature, exact dimer-dimer correlation functions of CDM are calculated in ref\cite{PhysRev.132.1411}, ref\cite{doi:10.1063/1.1703953} and ref\cite{PhysRev.124.1664}. The decay of dimer-dimer correlation function is algebraic, $G(x)\sim x^{-\alpha}$ where $x$ is the distance of two dimers. For example, horizontal dimers along the $x$-axis, the exponent $\alpha=2$.  Horizontal dimers along the $y$-axis, the exponent $\alpha=2$ if $x$ is odd and $\alpha=4$ if $x$ is even. Researchers studied the decay is still algebraic in ECDM covering fractional $N4$ dimers in ref\cite{PhysRevB.73.144504}, but we still do not understand the change of the exponent $\alpha$ for the case of larger fugacity.

\section{\label{sec:level1}NUMERIC RESULTS}
Firstly, we set a series of weight fugacity to study the evolution of order parameter at finite temperatures (lattice scale $L\times L=12\times 12$). At low temperatures, dimer symmetry breaks spontaneously. States are frozen at columnar crystal so the value of $COL$ of all curves saturates to 1 at the beginning. As temperature increases, dimer crystal melts to the liquid or gases so that the modules of columnar order parameters decrease and approach zero in the thermodynamic limit. For traditional CDM (in our model, it is the case of $f=0$), the phase transition of $COL$ is a KT phase transition, and the values of $COL$ are higher than zero.\cite{PhysRevLett.94.235702,PhysRevE.74.041124}

\begin{figure*}[htb]
	\centering
	\label{2}  
	\includegraphics[scale=0.8]{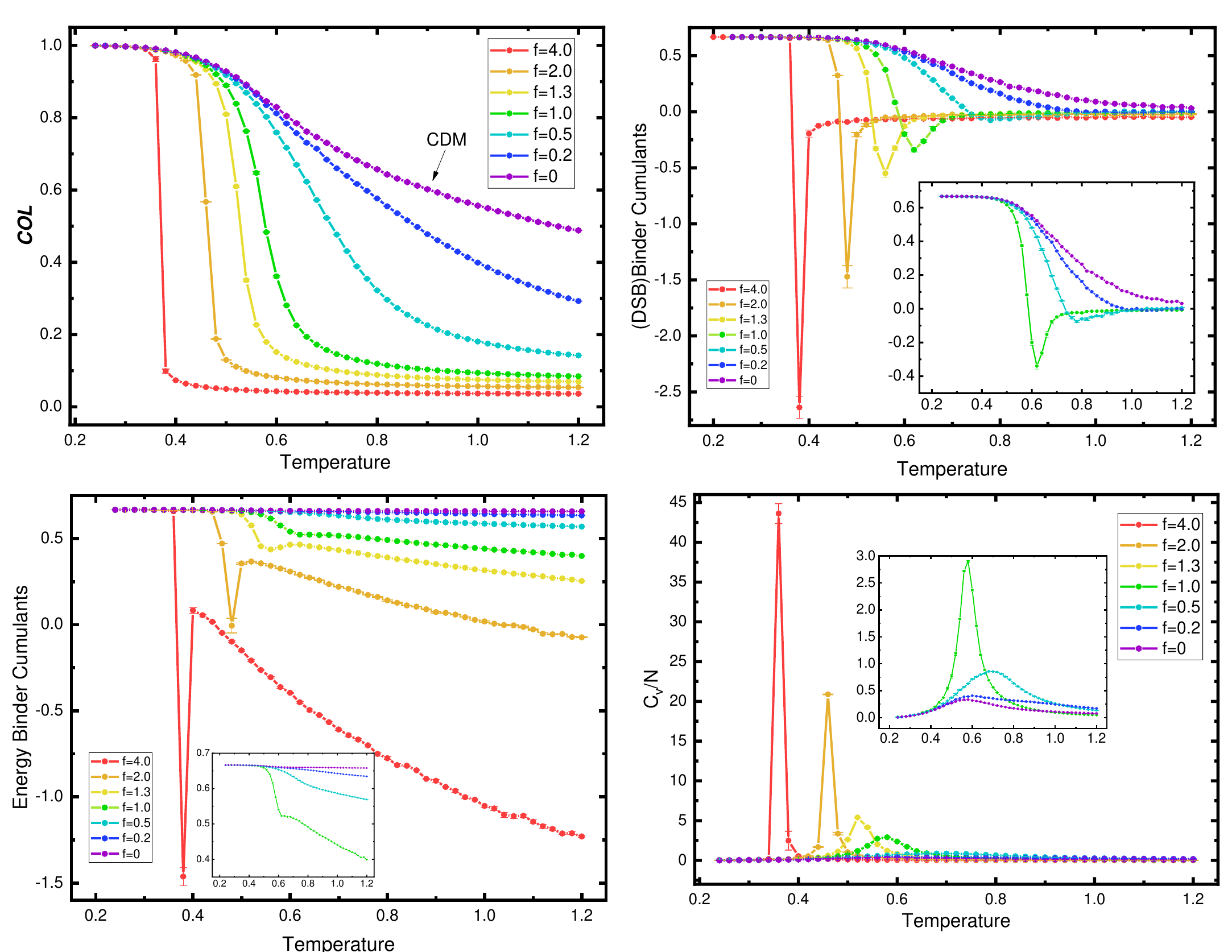}
	\caption{(Color online) (a). Complex order parameter $COL$ versus temperature $T$ for different weight fugacity. (b). Dimer rotation symmetry breaking ($DSB$) Binder cumulants versus temperature for different weight fugacity. Inset: Zoom-in figure of four small weight ($f=1.0$, $f=0.5$, $f=0.2$ and $f=0$) fugacity indicates the missing of negative peaks as fugacity decreases. (c). Energy cumulant versus temperature for different weight fugacity. Inset: Zoom-in figure of four small weight ($f=1.0$, $f=0.5$, $f=0.2$ and $f=0$) fugacity indicates the missing of negative peaks as fugacity decreases. (d). Specific heat per site $C_V/N$ versus temperature $T$ for different weight fugacity. The size of our simulation here is $L=12$. comparing with curves of $f=4.0$, curves of small weight fugacity loses their resolution to find the peak. Inset: Zoom-in figure of four small weight ($f=1.0$, $f=0.5$, $f=0.2$ and $f=0$) fugacity indicates the missing of peaks as fugacity decreases.  The size in our simulation here is $L=12$. Purple line respond to CDM and its weight fugacity $f=0$. The purple line decays to zero until $T=2.0$ in ref\cite{PhysRevE.74.041124} not shown in this figure.}
\end{figure*}

As we increase the fugacity, the decay of $COL$ curves tends to be sharp. For $f=0$, confinement of dimers requires the phase transition is mild reflected by $COL$ curves. However, as $N4$ dimers enter the dimer configuration, strong geometric constraints of dimers break up. The phase transition is not milt anymore and we show the steep decays appear when $f>0.5$. $COL$ curves give signals of the type of the phase transition reduced from KT transition to finite order phase transition.	
	
Considering the rotation symmetry of dimers, the fugacity dependence of Binder cumulant also displays abnormal evolution. $DSB$ Binder cumulant is defined as\begin{equation}
	B_{D} = 1-\left \langle DSB^4  \right \rangle/(3\left \langle DSB^2  \right \rangle^2)
\end{equation}
From the perspective of energy, we define the heat capacity per site as the numeric derivative and energy Binder cumulants as the fourth-order cumulants of energy,\begin{equation}
	\frac{C_v}{N}=\frac{\left \langle E^2  \right \rangle-\left \langle E  \right \rangle^2}{NT^2}
\end{equation}
\begin{equation}
	V=1-\frac{\left \langle E^4  \right \rangle}{3\left \langle E^2  \right \rangle^2}
\end{equation}  
For $f = 0$, there are peakless curve for Binder cumulant of $DSB$ and Binder cumulants of energy. Binder cumulants reflect the information of Gaussian distribution of the statistic quantities. All curves also saturate to $2/3$ at $T=0$. For ordered dimer states of columnar crystal, order parameter and energy have no fluctuations. In the thermodynamic limit, both of order parameter and energy stay Gaussian distribution with strong thermal fluctuations, so Binder cumulants converge to zero for these disorder states. Novel points appear when the system allows two or more states at a fixed point, and the distribution is a superposition of some Gaussian peak. This indicates that there are negative peaks in the curves of Binder cumulants at finite temperatures. For traditional CDM, curves cross over at the phase transition point and smoothly arrive at zero at high temperatures without any peaks\cite{Finiteesize}. As fugacity increases, negative peaks appear and the values of peaks can be enlarged. Another piece of evidence is the heat capacity of the system. For KT transition in CDM, heat capacity is suppressed under the value of $0.5$, and values of peaks are not enlarged as the sizes of the system increase. But for our ECDM, if the fugacity $f\gtrsim1$, maximums of heat capacity are diverging as the sizes of the system increase. We argue that the phase transition is reduced if we introduce $N4$ dimers. Continuity of the phases transition could be broken. Also, from the breaking of the continuous KT phase transition, the quasi-long order of the states could be broken simultaneously. We will demonstrate these arguments in the following. 

Next, we choose three characteristic fugacity to compare the detailed properties of phase transitions. $f=0.1$, $f=1.0$ and $f=2.0$. For $f=0.1$, fractional $N4$ dimers are admitted to enter the fully compact dimer model. The phase transition is still a KT transition with the same properties as the CDM. For $f=2.0$, large numbers of $N4$ dimers are allowed to enter the lattice after the critical points. At the critical temperatures, the states can be occupied by ordered columnar states and disorder $N4$ states. It means the phase transition is reduced to the first-order phases transition. For the meddle case of $f=1.0$, $N4$ dimers can impact the phase transition. The breaking of geometric constraints of the dimer formation influences the quasi-ordered range of states after critical points. At this time, the phase transition reduces from a KT transition to a high-order phase transition with the properties of the first-order phase transition and KT transition. We compare free holes of the dimer-hole model to analyze the effect of $N4$ geometric constraints in ECDM.

\begin{figure*}[htb]
	\centering
	\label{3} 
	\includegraphics[scale=0.24]{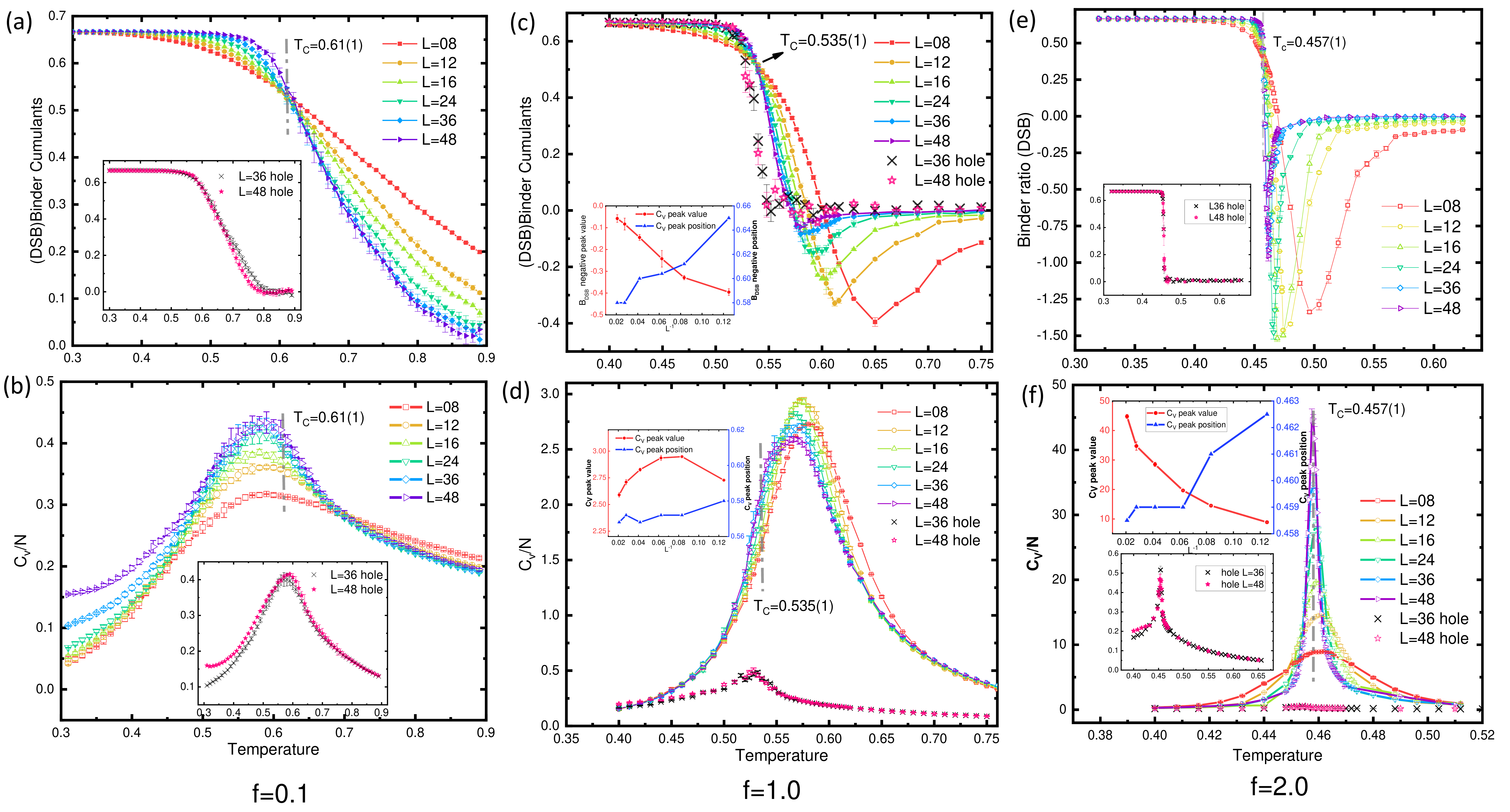}
	\caption{(Color online) (a)(b). The scale dependence of dimer rotation symmetry breaking ($DSB$) Binder cumulants and specific heat per site $C_V/N$ versus temperature for $f=0.1$.  The dashed line indicates $T_c=0.61(1)$ identified by the cross-over point of dimer rotation symmetry breaking ($DSB$) Binder cumulants curves. Inset: $DSB$ Binder cumulants in Fig. 3(a) and specific heat per site $C_V/N$ in Fig. 3(b) versus temperature of the system simulated with holes of same values of $N4$ dimers corresponding to the system for $f=0.1$. (c)(d). The scale dependence of $DSB$ Binder cumulants and specific heat per site $C_V/N$ versus temperature for $f=1.0$. Inset: values and position of negative peaks in $DSB$ Binder cumulants curves versus $L^{-1}$ and values and position of peaks in Heat capacity curves versus $L^{-1}$. (e)(f). The scale dependence of $DSB$ Binder cumulants and specific heat per site $C_V/N$ versus temperature for $f=2.0$. Inset: values and position of negative peaks in $DSB$ Binder cumulants curves versus $L^{-1}$ and values and position of peaks in Heat capacity curves versus $L^{-1}$. Inset: $DSB$ Binder cumulants in Fig. 3(e) and specific heat per site $C_V/N$ in Fig. 3(f) versus temperature of the system simulated with holes of same values of $N4$ dimers corresponding to the system for $f=2.0$.  In Fig. 3(f), upper figure shows values and position of $C_V/N$ curves peaks.
	}
\end{figure*}
	
In Fig. 3(a) and 3(b), the properties of $DSB$ Binder cumulants and heat capacity are the same as the CDM\cite{PhysRevLett.94.235702,PhysRevE.74.041124}. Curves of DSB Binder cumulants approach zero smoothly and these curves cross over at the critical point $T=0.61(1)$. We consider the same holes in the dimer-hole model in the inset of Fig. 3(a) and 3(b). $DSB$ Binder cumulants and heat capacity of the dimer-hole model are similar to the ECDM. It indicates that the influence of $N4$ dimers is not unique for a small fugacity. Fractional $N4$ dimers are distributed rarely in the ECDM and they have no power to form the effect to destroy the quasi-long range order after the critical point.

For $f=2.0$, DSB Binder cumulants curves also cross over at a point that identifies the critical point as $T\sim0.457(1)$ in Fig. 3(e). The peak of heat capacity per site is located at $T_C\sim0.457(1)$, denoted by the dashed line in Fig. 3 (f). Both heat capacity and $DSB$ binder cumulants identify the same critical point without shift. This is a difference between the previous KT phase transition. Another difference between $f=2.0$ and $f=0.1$ is there are negative peaks in the curves of $DSB$ Binder cumulants. The values and position of these peaks are the size dependence. The width and the height of the size $L=8$ are wider and higher than others due to the effect of finite size. As we increase the size of the ECDM, the values of the negative peak reduce smaller, and the broadenings of the peaks narrow. The broadening is narrow, so we are difficult to search peaks of higher precision for large size. In our simulation, values of peaks for $L=36$ and $L=48$ are approaching and they do not reduce any more. The values of peaks for heat capacity are much larger than CDM and we show the values of peaks are amplified by system size shown in the inset of Fig. 3(f). There is no shift of these peaks as the size increases. the properties of negative peaks in the $DSB$ Binder cumulants curves and the diverging peaks of heat capacity imply the phase transition is the first-order phase transition here. 

\begin{figure*}[htbp]
	\centering
	\label{4} 
	\includegraphics[scale=0.53]{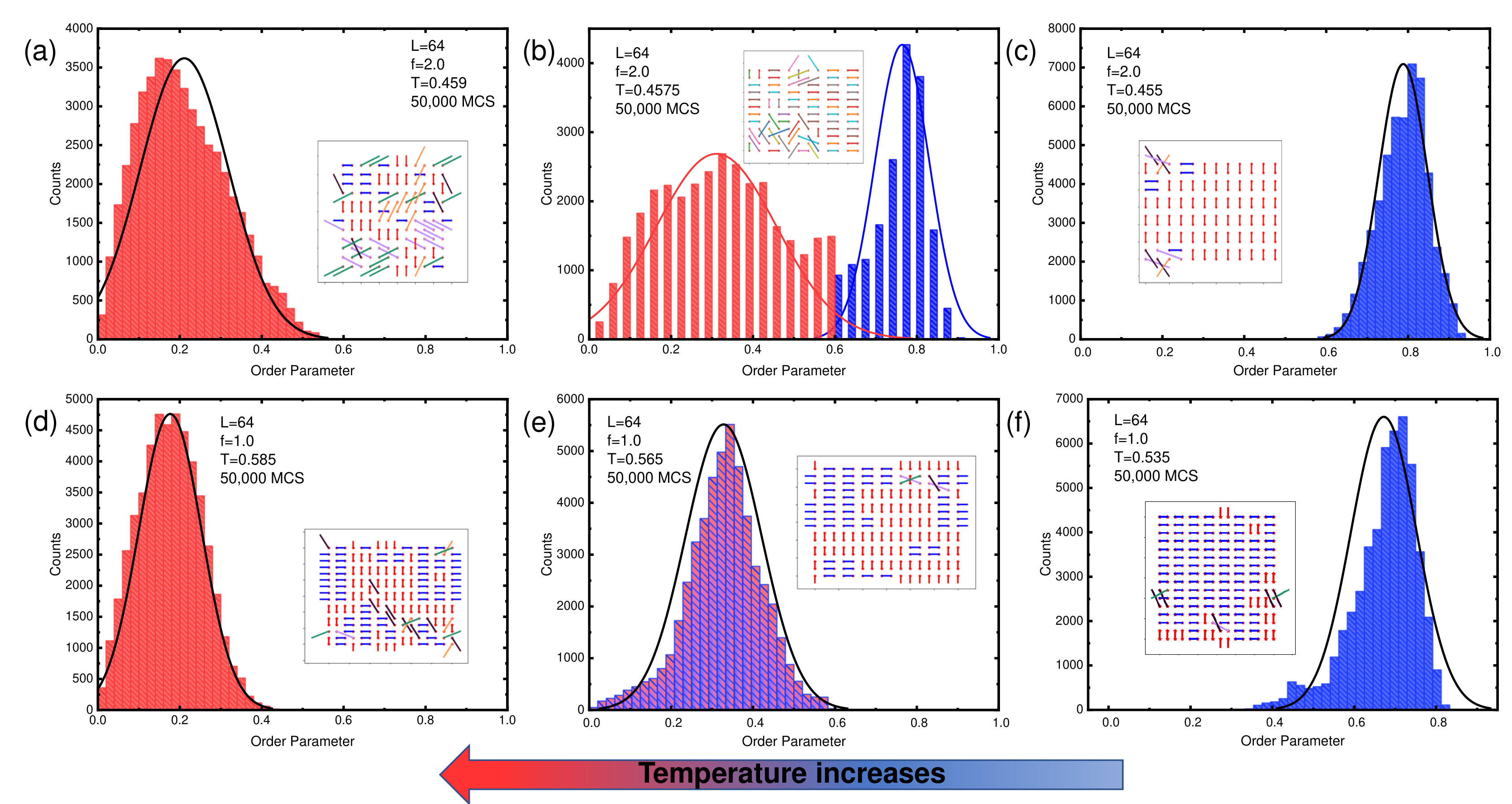}
	\caption{$COL$ order parameter distribution at different temperatures around the point of phase transition. (a)(b)(c). the evolution of double peaks and the superposition of two Gaussian distribution of the ECDM for $f=2.0$ as temperatures increase. (d)(e)(f). single peak and Gaussian distribution of the ECDM for $f=1$.  Inset: schematic dimer configurations of characteristic states simulated in the size of $L=16$ at corresponding temperatures.  
	}
\end{figure*}

For $f=1.0$, this point is a characteristic point chosen by us to study the phase transition between the above two fugacity. In Fig. 2(a), COL shows no obvious sharp decay for $f=1.0$. Peaks of heat capacity do not distribute along a vertical line and we can just identify the phase transition point as $T_c=0.56(1)$ approximately. The cross-over phenomenon of Binder cumulants identifies the transition point as $T_c=0.535(1)$. However, there are some strange phenomena in $DSB$ Binder cumulants and heat capacity curves. For the $DSB$ Binder ratio, negative peaks appear in curves of Binder cumulants shown in Fig. 3(c). These peaks decrease as the size increases. In the inset, we show the curves of values and the position of negative peaks versus $L^{-1}$. The negative peaks can be extrapolated to zero yielding that the phase transition is not the first-order phase transition. The values of peaks for heat capacity shrink with a small trend as the system size enlarges shown in Fig. 3(d). In the scale of $L=8$, the peak is not the highest due to the finite scale effect. These converge peaks are similar to the heat capacity of CDM. However, the values of peaks for $f=1.0$ are larger than CDM and they are larger than $T_c=0.535$. In contrast, in the original CDM the peak of heat capacity is significantly lower than the transition temperature $T_c$. It indicates $N4$ dimers can introduce larger thermal excitation to give rise to larger energy change than the normal excitation of $N1$ dimers. The phenomenon is observed in CDM but not in the case of $f = 2.0$ ECDM. The phase transition excludes the first-order phase transition but we still need more evidence to justify if the phase transition is a KT transition for $f=1.0$. 

We count the number of $N4$ dimers in the system of $f=1.0$ and $f=2.0$ and simulate the dimer-hole model with the same values of holes. $DSB$ Binder cumulants and heat capacity show that there are no negative peaks and diverging peaks in their curves for both of the cases. Excitations of holes generate lower energy change than $N4$ thermal excitations. It indicates the effect of the geometric constraints in $N4$ dimers.

Furthermore, to confirm the type of phase transitions for $f=1.0$ and $f=2.0$, we plot the distribution of order parameters at a definite temperature shown in Fig. 4. We show evidence of phase coexistence, which is a symbolic feature of the first-order phase transition for $f=2.0$. Otherwise, there is no evidence to manifest the phase transition as the first-order phase transition for $f=1.0$. We simulate the model with the size of $L=64$. Fundamentally, the distribution of thermodynamic fluctuations is Gaussian. However, the characteristic feature of first-order phase transition is phase coexistence at the transition point in ref\cite{PhysRevB.34.1841}. We show the probability  $P_L(COL)$ distribution of COL instead of energy like the ref\cite{PhysRevB.34.1841}. For the first-order phase transition, $P_L(COL)$ is a superposition of two Gaussian distribution. Two distributions are centered at $COL_{+}$ and $COL_{-}$. As a simple schematic illustration of phase coexistence, we sample states at corresponding temperatures of $L=16$ to show features of states corresponding to every temperature in insects of Fig. 4. Obviously, below the transition temperature, dimers are still assigned ordered as low-temperature crystals. There are only seldom disordered dimers, so COL distributes Gaussian around the point approximating to 1 shown in Fig. 4(a). As temperature increases, the disordered area is expanded on the lattice. Approximating to the critical point, the configuration is occupied by ordered dimers and disordered dimers simultaneously shown in Fig. 4(b). After the critical point, the distribution of COL recovers to the normal Gaussian without superpositions shown in Fig. 4(c). For $f = 2$, we can confirm that the phase transition is reduced to the first-order phase transition. But for the case of $f=1.0$, the result of the distribution is trivial Gaussian distribution like CDM in Fig. 6. There is no superposition of two Gaussian distributing at the critical point. We can exclude the transition is the first-order phase transition compared with the case of $f=2.0$ shown in Fig. 4(d)-(f).

\begin{figure*}[htb]
	\centering
	\label{5} 
	\includegraphics[scale=0.74]{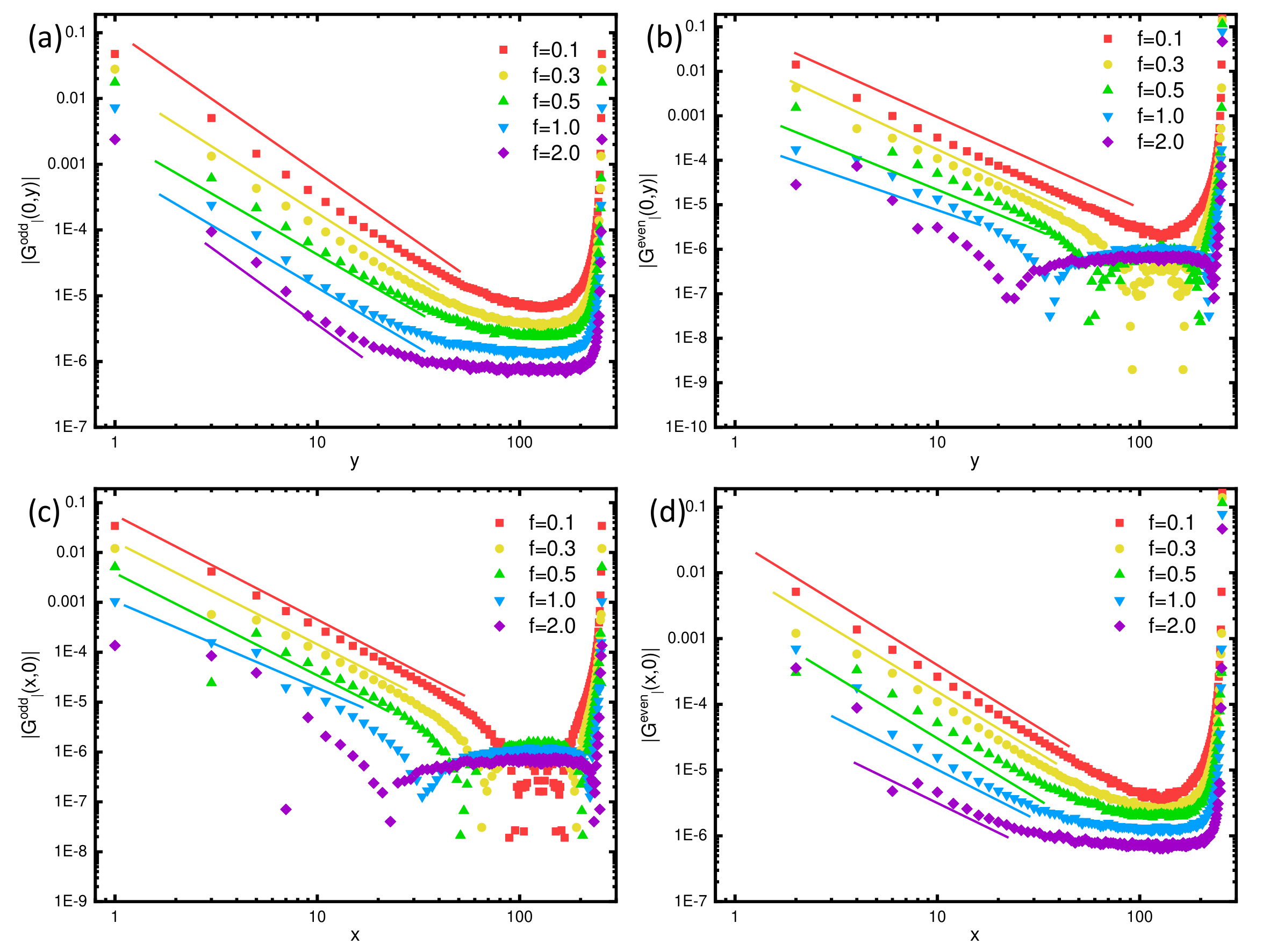}
	\caption{(Color online) Dimer-dimer correlation function $G_{\alpha}^{\beta}(x,y)$ versus vectors $(x,y)$ for different fugacity at $T=\infty $ on a log-log scale. (a)(b): correlation function $G_{\alpha}^{\beta}(x,y)$  of vertical dimers along vertical direction for odd position of $y$ and even position of $y$. (c)(d): correlation function $G_{\alpha}^{\beta}(x,y)$  of vertical dimers along horizontal direction for odd position of $x$ and even position of $x$. Lines indicate linear distribution of part of scatters in a fixed fugacity schematically.
	}		
\end{figure*}

We compute the dimer-dimer correlation functions and their structure factors to study the order of high-temperature states. We define correlation functions in the following:\begin{equation}
	G_d(x,y)=\left \langle \hat{n}_d(\textbf{r})\hat{n}_d(\textbf{r}+\textbf{u}) \right \rangle_r-\left \langle \hat{n}_d(\textbf{r})\right \rangle_r\left \langle \hat{n}_d(\textbf{r}+\textbf{u}) \right \rangle_r
\end{equation}

Where $\textbf{u}=(x,y)$ and both $x$ and $y$ are integers.  $\hat{n}_d(\textbf{r})=1$ for the corresponding dimer at site $\textbf{r}$, otherwise $\hat{n}_d(\textbf{r})=0$.

For CDM, after the KT phase transition, the system still retains its quasi-long-range order due to the confinement of rigid geometry. In this paper, we deconfine the condition of nearest bonding, and extend it to the second-nearest-neighbor bonding. In early research, M. E. Fisher et al. computed the dimer-dimer correlation function by the method of Kasteleyn's theorem as ref\cite{PhysRev.132.1411,doi:10.1063/1.1703953,PhysRev.124.1664}.\begin{equation}
	G_-(x,0)\sim\frac{(-1)^x}{\pi^2x^2}+O(x^{-3})
\end{equation}
And\begin{equation}
	G_-(0,y)\sim\frac{1}{\pi^2y^2}+O(y^{-3}),\quad y \quad are \quad odd
\end{equation}
\begin{equation}
	\sim-\frac{1}{\pi^2y^4}+O(y^{-6}),\quad y \quad are \quad even
\end{equation}

For dimer-dimer correlation functions, positive values mean there is correlation tending to form dimer at site $\textbf{r}+\textbf{u}$ if dimer exists at site $\textbf{r}$ and negative values mean there is reverse correlation tending to exclude dimer at site $\textbf{r}+\textbf{u}$ if dimer exits at site $\textbf{r}$. For the easiest case, if a horizontal dimer exists at site $\textbf{r}$, there must be no dimers at site $\textbf{r}+(1,0)$. Also, at low temperatures, if a horizontal dimer exists at site $\textbf{r}$, interaction drives dimers to be parallel to it, so there are dimers at site $\textbf{r}+(0,y)$. 

The algebraic decay is dominated by the term of exponent $\alpha=2$for $G_{-}(x,0)$, and exponent $\alpha=2$ and $\alpha=4$ for $G_{-}(0,y)$. It means there is a quasi-long-range order in the thermodynamic limit. We separate the odd terms and even terms for $G_{|}(0,y)$ and  $G_{|}(x,0)$ and show the dimer-dimer correlation functions on a log-log scale (lattice size $L=256$) in Fig. 5.

For odd terms of $G_{|}(0,y)$, when $f=0.1$, the correlation function retains the linear property and the exponent approaches $\alpha=2$. However, if we increase the fugacity, linearity only exists at the front points and exponents go down. For $f=2.0$, a flattened plat appears in the middle, and only $22$ points retain linearity. The platform is not on zero completely because of the effect of finite size. This evidence demonstrates that the quasi-long-range order breaks down and the length of the order is shortened due to the influences of $N4$ dimers. For even terms of $G_{|}(0,y)$, we can derive the same conclusions and there is a peak at $y=22$. Other correlation functions are consistent with the $G_{|}(0,y)$. The loss of linearity means that dominant terms of the correlation function are no longer $-2$. And the peak in even terms of $G_{|}(0,y)$ can be given by some exponents working together. Dimer-dimer correlation functions give a picture of the change of the quasi-long-range order in ECDM as the fugacity increases.

\section{\label{sec:level1}CONCLUSION}	
We introduce $N4$ dimers into interacting close-packed dimer model on square lattice at finite temperatures and study the phase transition process of different fugacity of $N4$ dimers as ECDM. We simulate the thermodynamic quantities and types of order parameter to probe the phase transition and calculate the dimer-dimer correlation functions in any cases. We systematically study the properties in $k$-space of correlation functions. Among all thermodynamic quantities and correlation function can reveal a clear evolution of phase transition as we tune the fugacity. We derive a schematic phase diagram of our system shown in Fig. 6. Due to our constrained computational power, we do not sweep parameters in very small steps but describe the physics in each characteristic zone. 

For low fugacity ($f\ll0.5$), features of ECDM are approaching to CDM as a KT phase transition. For middle zone of weight fugacity ($0.5\lesssim f\lesssim1.25$), there are features of first-order phase transition and KT transition at the same time. As the increase of size of model, the properties of first-order phase transition can be shortened but the system still retain some difference from a pure KT transition. Our argument is the phase transition here retracts into a high-order phase transition beyond convention. For high fugacity zone ($f\gtrsim1.25$), the phase transition is a quasi-first-order phase transition due to the competition in configurations of $N1$ dimers and $N4$ dimers. Correlation functions indicates the breaking of quasi-long-range order.

We still hope to study the model with larger size to shorten the impact of finite-size effect. If $N4$ dimers own intrinsic energy or they can bond with each other, the system may face new competition and emerge new physics. In addition, the study of $N4$ dimers can offer a type new quantum fluctuation in QDM as quantum entanglement or defects impeding the formation of bound states of electrons. 

\begin{figure}[htb]
	\centering
	\label{6} 
	\includegraphics[scale=0.32]{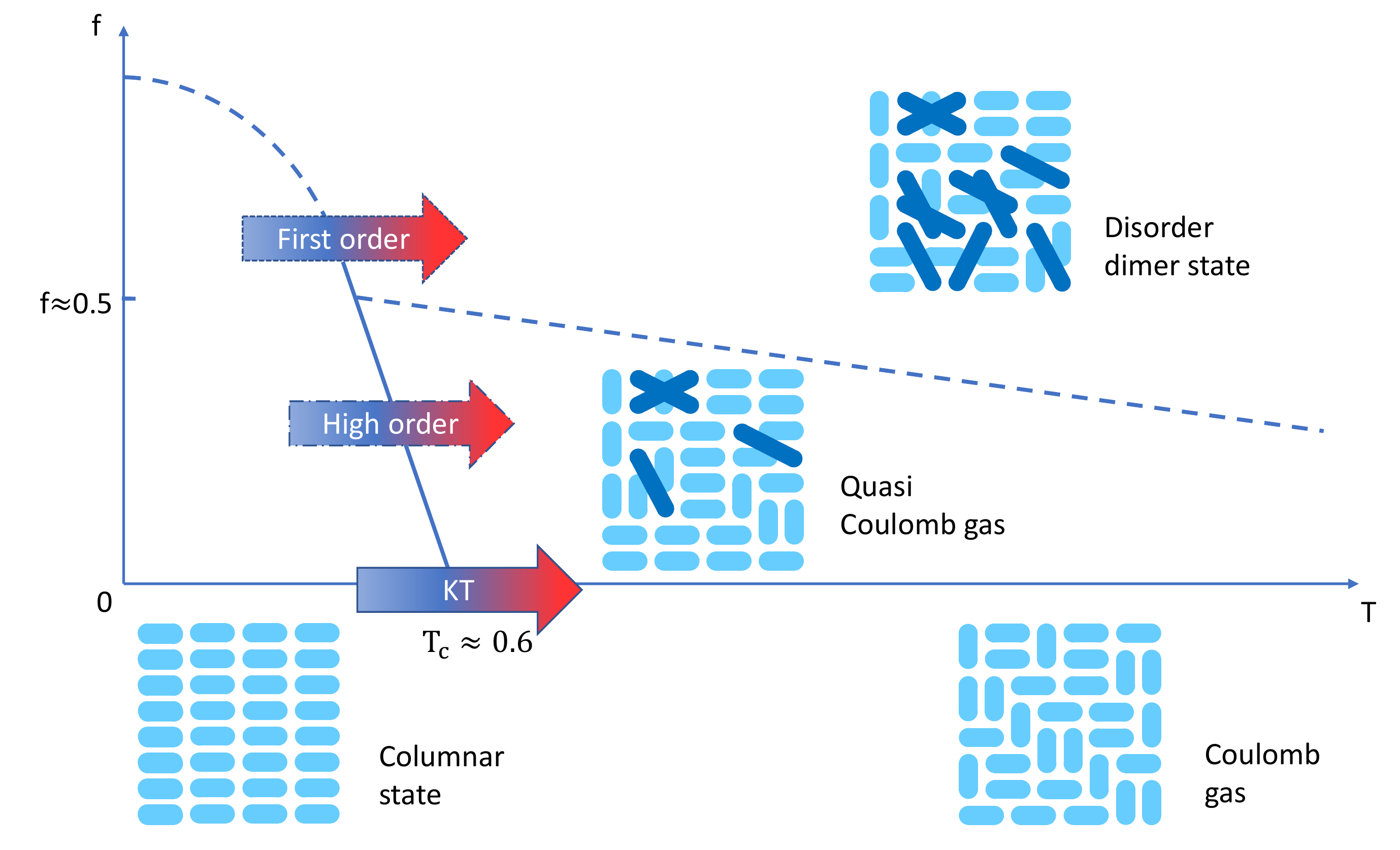}
	\caption{.}	
\end{figure}
	
\appendix
	
\section{ENERGY DIRECTED LOOP ALGORITHM}\label{app1}
Directed loop algorithm is introduced by A. W. Sandvik et al in 1/2 antiferromagnetic Heisenberg model firstly.\cite{doi:10.1063/1.1632141} They also introduce it to CDM and extend it as a universal algorithm for ECDM and many researchers use it to study CDM for a long venerable history.\cite{PhysRevLett.94.235702,PhysRevE.74.041124,PhysRevB.73.144504} The details of directed loop algorithm used by them are in their paper. Still, our usage of the directed loop algorithm in this paper is a little different but they are same in principle. We hope to introduce it in the following.
	
Recall that Monte Carlo simulation will be constructed under the fundamental of Metropolis sampling algorithm\cite{Computational}, we decompose the updating process into every steps of the generation of directed loops. 
	
The whole process is in the following:
	
1). We choose a vertex as our starting point and annihilate the occupied dimer here. After that, we define the other point of the dimer as a walking seed. In Fig. 7(a), we choose the blue point as the starting point and the red as our walking seed.
	
2). For the walking seed, there are four directions for it to create new dimer as the path shown in Fig. 7(b). It means we can create new dimer in four directions around the walking seed. The directed loop will step into one of these four directions as red arrows showing.
	
3). It is distinguishable to the regular directed loop, instead of just choosing one direction to walk randomly, we compute the change of energy of every path in four choices of direction and judge each of them by a prior Metropolis sampling. For example, in Fig. 7(c), we suppose the direction pointing to the top has been chosen. Thus, we compute the energy change if we create a new red dimer here. The red dimer can bond with the parallel dimer in the yellow dashed line frame. If the change of energy in the frame is accepted, we can derive a path of the directed loop. 
	
4). Then, we annihilate the cross-over dimer and repeat above steps until the energy-based directed loop comes back to the starting point and form a closed loop. In Fig. 7(d), the directed loop repeats above operations. 
	
If directed loops walk randomly, either the random walking forms local loops so that the system drops into sub-equilibrium states. Now, for energy-based directed loop algorithm, the addition of energy criterion induces a more efficient walking in the updating process in Monte Carlo simulation. Metropolis sampling judging in every steps of the generation of the energy-based directed loop depends the loop length and the change of energy matching with homologous thermal states. 
	
In the ECDM, detailed balance condition is\begin{equation}
	a_{sl}=a_{ls}
\end{equation}
We can see for a fixed weight fugacity $f$, there are infinite pairs of $\omega_l$ and $\omega_s$ satisfying the condition. For directed loop algorithm of random walking, for every pair of $\omega_l$ and $\omega_s$, only if do we fix the fugacity $f$, all simulation results are same. In other words, $a_{sl}=a_{ls}$ is sufficient and necessary for traditional directed loop algorithm. However, for energy-based directed loop algorithm, we can not derive same results now. This condition is not sufficient now. The choice of different   $\omega_s$ and $\omega_l$ impact the probability of directions of every path.
	
Our target is to achieve random update but fixed fugacity, so we need totally random direction of the energy path. From $a_{sl}=a_{ls}$, we can re-write the weight of $N1$ dimers and $N4$ dimers.\begin{equation}
	\omega_s^{'} =\frac{\omega_{ss}}{a_{sl}} =\frac{4a_{ss}}{a_{sl}}+8
\end{equation}
\begin{equation}
	\omega_l^{'} =\frac{\omega_{ll}}{a_{ls}} =\frac{8a_{ll}}{a_{ls}}+4
\end{equation}
Hence, the probability of annihilating an old $N1(N4$) dimer and creating a new $N1(N4)$ dimer is\begin{equation}
	P_{ss}=\frac{4a_{ss}}{\omega_s^{'} } =\frac{\omega_s^{'} -8}{\omega_s^{'}}
\end{equation}
\begin{equation}
	P_{ll}=\frac{8_{ll}}{\omega_l^{'} } =\frac{\omega_l^{'} -4}{\omega_l^{'}}
\end{equation}
To satisfy the condition of random directions, we let\begin{equation}
	P_{ss}=P_{ls}
\end{equation}
and
\begin{equation}
	P_{ll}=P_{sl}
\end{equation}
Finally, we can derive\begin{equation}
	\frac{8}{\omega_s} +\frac{4}{\omega_l}=1
\end{equation}
Consider the second equation, for a fixed fugacity $f$, the detailed balance condition can be satisfied for energy-based directed loop algorithm in ECDM for only a pair of $\omega_s$ and $\omega_l$. 
\begin{figure}[htb]
	\centering
	\subfigure[]{ \label{7a} \includegraphics[width = 3.3cm, height = 3.3cm]{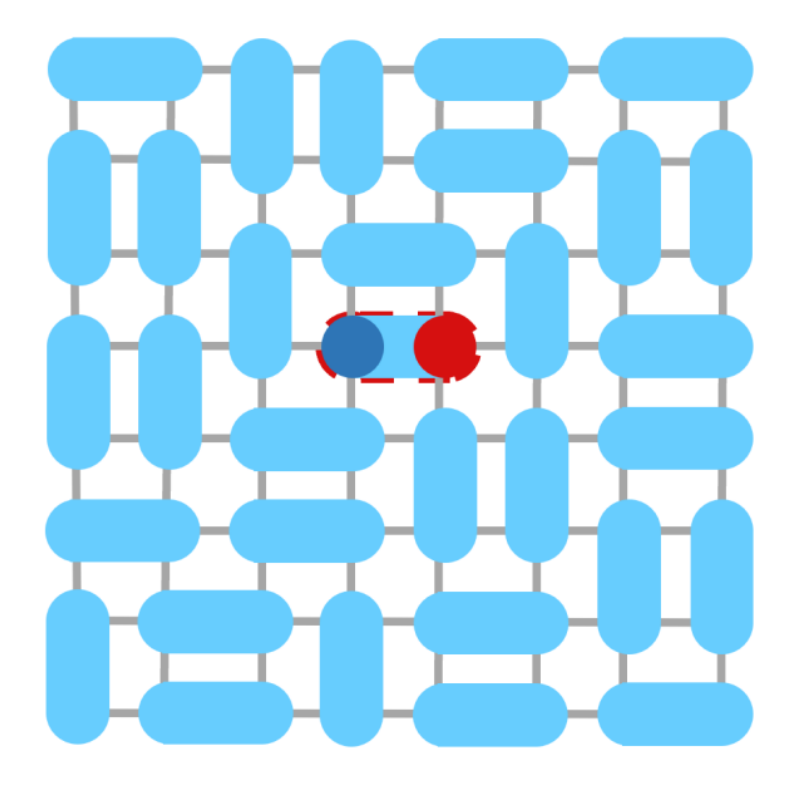}}
	\subfigure[]{ \label{7b} \includegraphics[width = 3.3cm, height = 3.3cm]{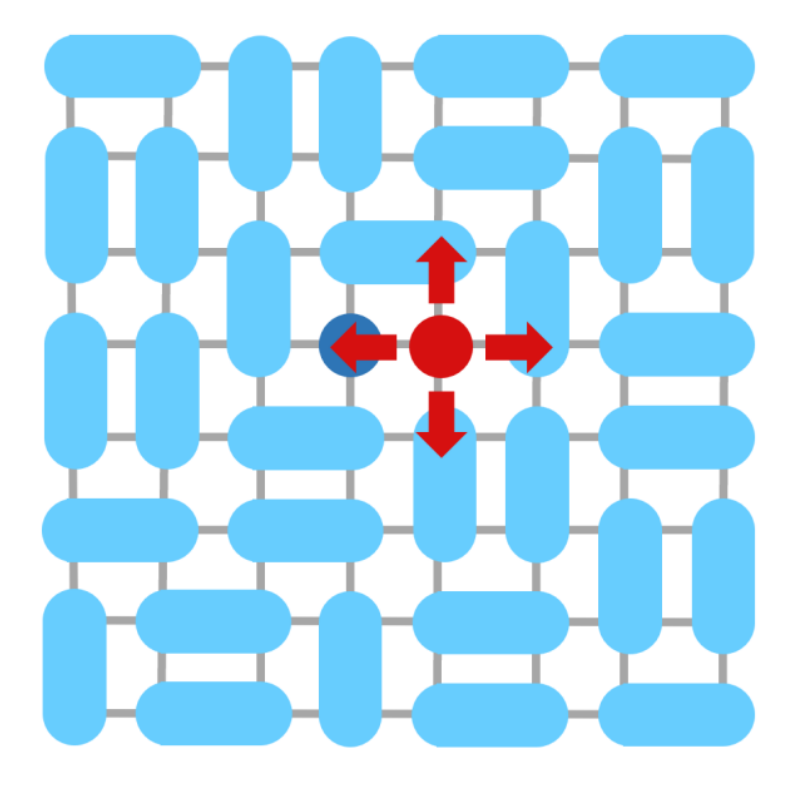}}
	\subfigure[]{ \label{7c} \includegraphics[width = 3.3cm, height = 3.3cm]{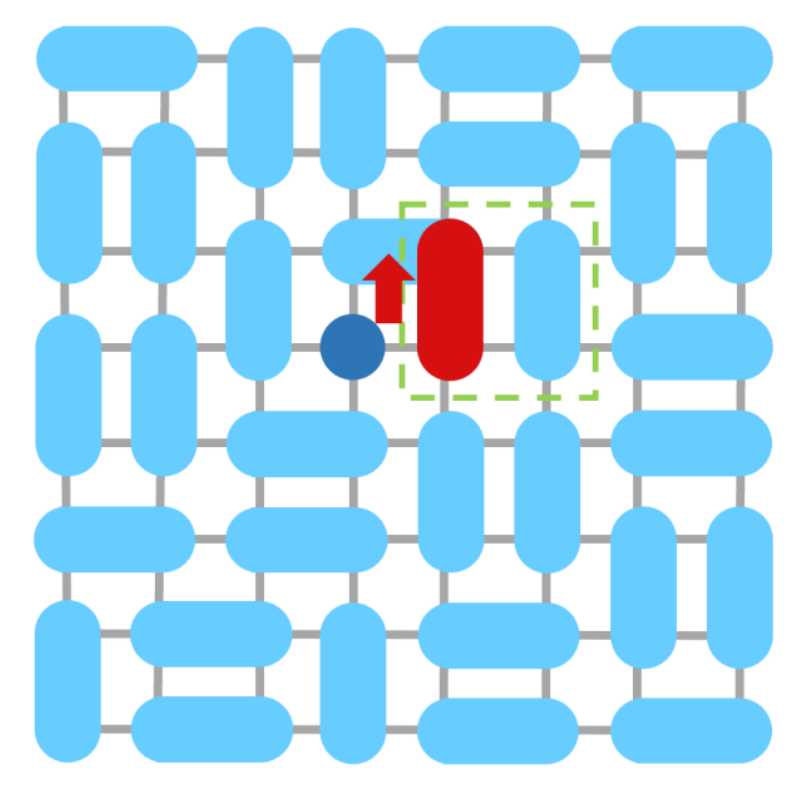}}
	\subfigure[]{ \label{7d} \includegraphics[width = 3.3cm, height = 3.3cm]{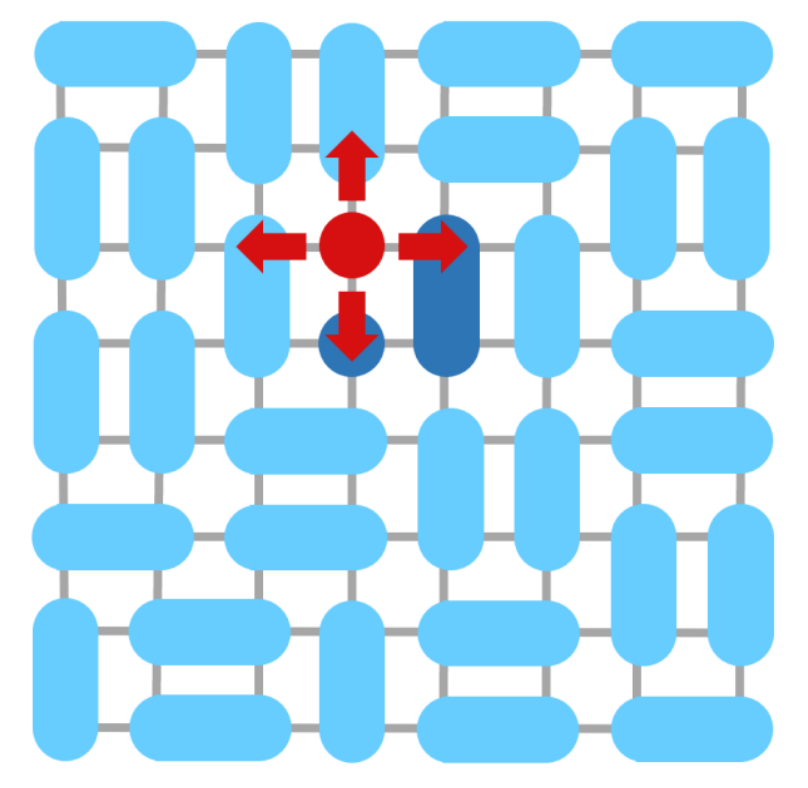}}
	\caption{(a)(b)(c)(d): schematic process of eDLA(blue point is the origin of the upcoming path and the red point is the directed point under energy criterion. Red dimer is the judging dimer and the blue dimer is the chosen dimer). }		
\end{figure}

\bibliography{dimersarticle2}%Produces the bibliography via BibTeX.
	
\end{document}